\documentclass{kluwer}

\usepackage{graphicx}

\newcommand{\ra}{$\rightarrow$}

\begin{document}

\begin{article}

\begin{opening}

\title{The evolution of black hole states}

\author{Jeroen \surname{Homan}}
\institute{MIT Center for Space Research\\
70 Vassar Street, Cambridge, MA 02139, USA\\
jeroen@space.mit.edu}
\author{Tomaso \surname{Belloni}}
\institute{INAF/Osservatorio astronomico di Brera\\
Via E.~Bianchi 46, 23807, Merate (LC), Italy\\
belloni@merate.mi.astro.it}

\begin{abstract} We discuss the evolution of black hole transients on
the basis of a few systems that were intensively observed with the
{\it Rossi X-ray Timing Explorer}. We focus on the global evolution
and the observed state transitions. Rather than giving a numerical recipe for classifying observations, we try to identify times during outbursts at which clear changes occur in the X-ray variability, X-ray spectral, or multi-wavelength properties. 

\end{abstract}

\end{opening}

\section{Introduction}

Since the launch of the {\it Rossi X-ray Timing Explorer (RXTE)}
about $\sim$20 black hole X-ray transients have been observed with
enough coverage to study their global evolution. These observations
have provided a wealth of information and have already led to a
considerable increase in our understanding of these systems. In a
recent review by \inlinecite{mcre2003} a new classification scheme
was proposed for the spectral and variability states as observed in
black hole transients. \inlinecite{mcre2003} approached the states
issue by identifying three `stable' states in black hole transients.
In this paper we approach the issue from an other angle, focusing on
the observed transitions, the overall evolution during an outburst,
and how X-ray changes relate to changes at other wavelengths.

\section{State definitions}

Until the late 1990s it was generally assumed that the state of a
black-hole system was determined by the instantaneous mass accretion
rate, $\dot{M}$. It was believed that as the mass accretion rate
increased, a source went through the following states (see
e.g.~\opencite{esmcna1997}) : quiescence \ra\ low/hard \ra\
intermediate \ra\ high/soft \ra\ very high. Observations with {\it
RXTE} of sources like XTE J1550--564 have shown that such a simple
$\dot{M}$-driven picture, with transitions only being triggered by
changes in $\dot{M}$, is probably not able to explain the observed
transition between the states. \inlinecite{howiva2001} suggested that
an additional parameter may play a role in those transitions. They
also suggested that the very-high state and the intermediate state(s)
are one and the same state and represent transitions between the
low/hard and high/soft states at different luminosities. The fact
that most of the states were observed over wide and overlapping
ranges in luminosity also meant that attributes like `low', `high'
and `very high' had lost most of their significance.

\inlinecite{mcre2003} introduced a new classification scheme that is
partly based on the old five-states scheme, but no longer uses
luminosity as a selection criterion since it appeared that any of
the  active states may occur at any luminosity. They still recognize
a quiescent state, hard state, soft state (renaming the latter the
`thermal dominant state'), and very-high state (renaming it the steep
power-law state), but drop the intermediate state as a bona fide
state. The three active states are defined on the basis of the
fractional contribution of the disk-flux (or power-law flux) to the
2--20 keV spectrum, spectral power-law index, and the strength of the
power continuum. All observations that cannot be classified according
to their state definitions, are combined into an intermediate state,
in which observations can show properties of any of the three main
states. 

Some of the state transitions discussed in this paper involve (and
are defined on the basis of) sudden changes in the properties of the
quasi-periodic oscillations (QPOs) in the $\sim$1--10 Hz range.
Currently three types of these low-frequency QPOs are recognized in
black hole binaries, called type A, B and C (\opencite{wihova1999},
\opencite{resomu2002}). The three types can be distinguished on the
basis of strength, coherence, phase lags, energy dependence, harmonic
content, and frequency stability on a time scale of days. Type C QPOs
are the most common ones; they are observed in the spectrally hard
states over a wide range in frequency ($\sim$0.1--10 Hz), and are
stronger and more coherent than the other two types. Type A and B
QPOs have only been observed in a few sources; they are only seen in
the very-high/steep power-law state, in a narrow frequency range ($\sim$4--8 Hz). Recent
observations of H1743--322 (\opencite{homiwi2005}) suggest that these
two types might be more intimately related than was previously
believed. Transitions from one type of QPO to another always seem to
involve type B (\opencite{cabeho2004}), with transitions between type
C and B resulting in a clear change in the shape and strength of
noise continuum. The main reason for including QPO type in our
discussion of black hole states is that they provide an additional
indication for changes in the accretion flow that may not always show
up as strong spectral transitions.



\section{Global evolution}

\begin{figure}[!t]
\centerline{\includegraphics[width=7cm]{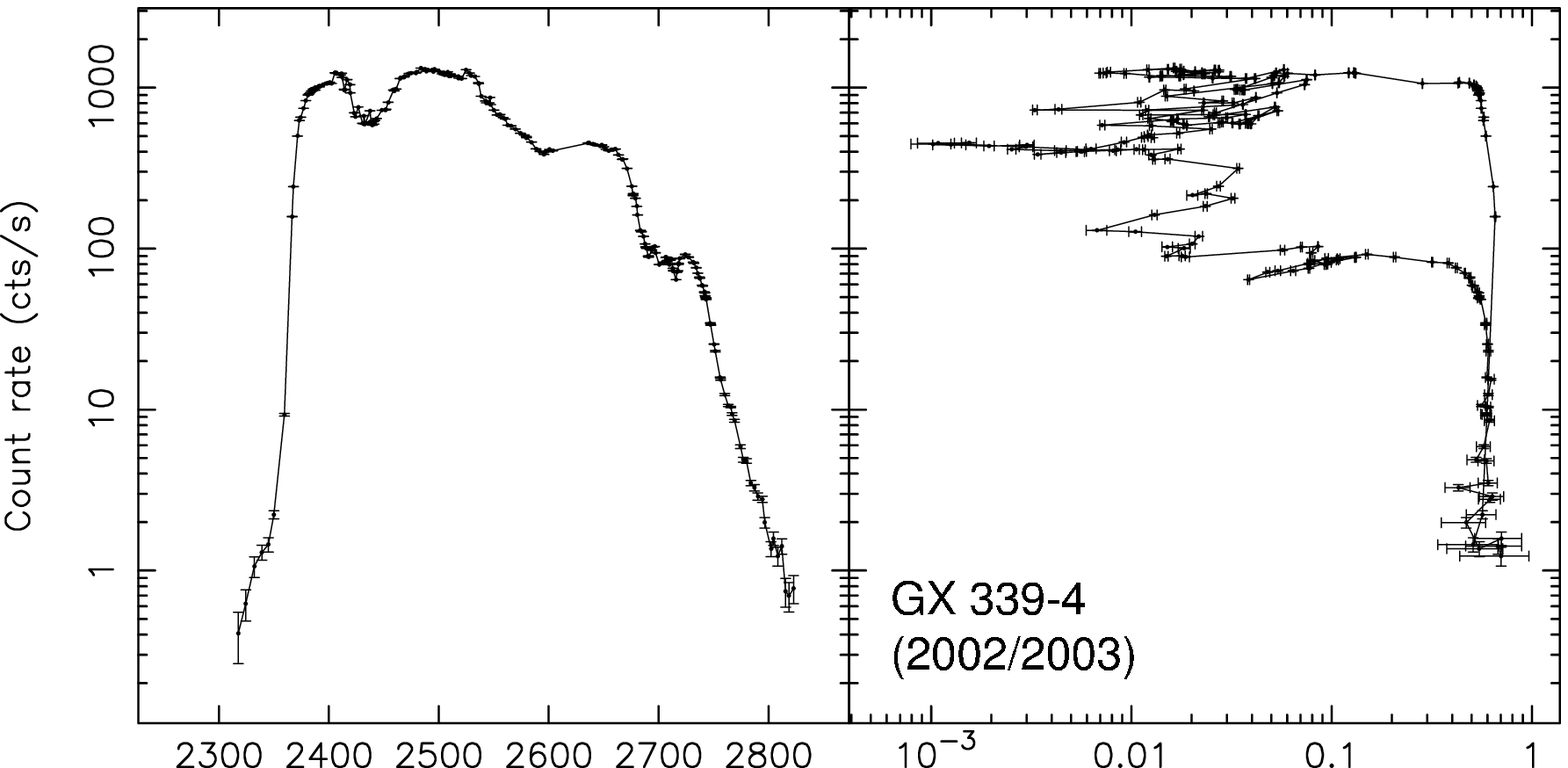}}
\vspace{0.1cm}
\centerline{\includegraphics[width=7cm]{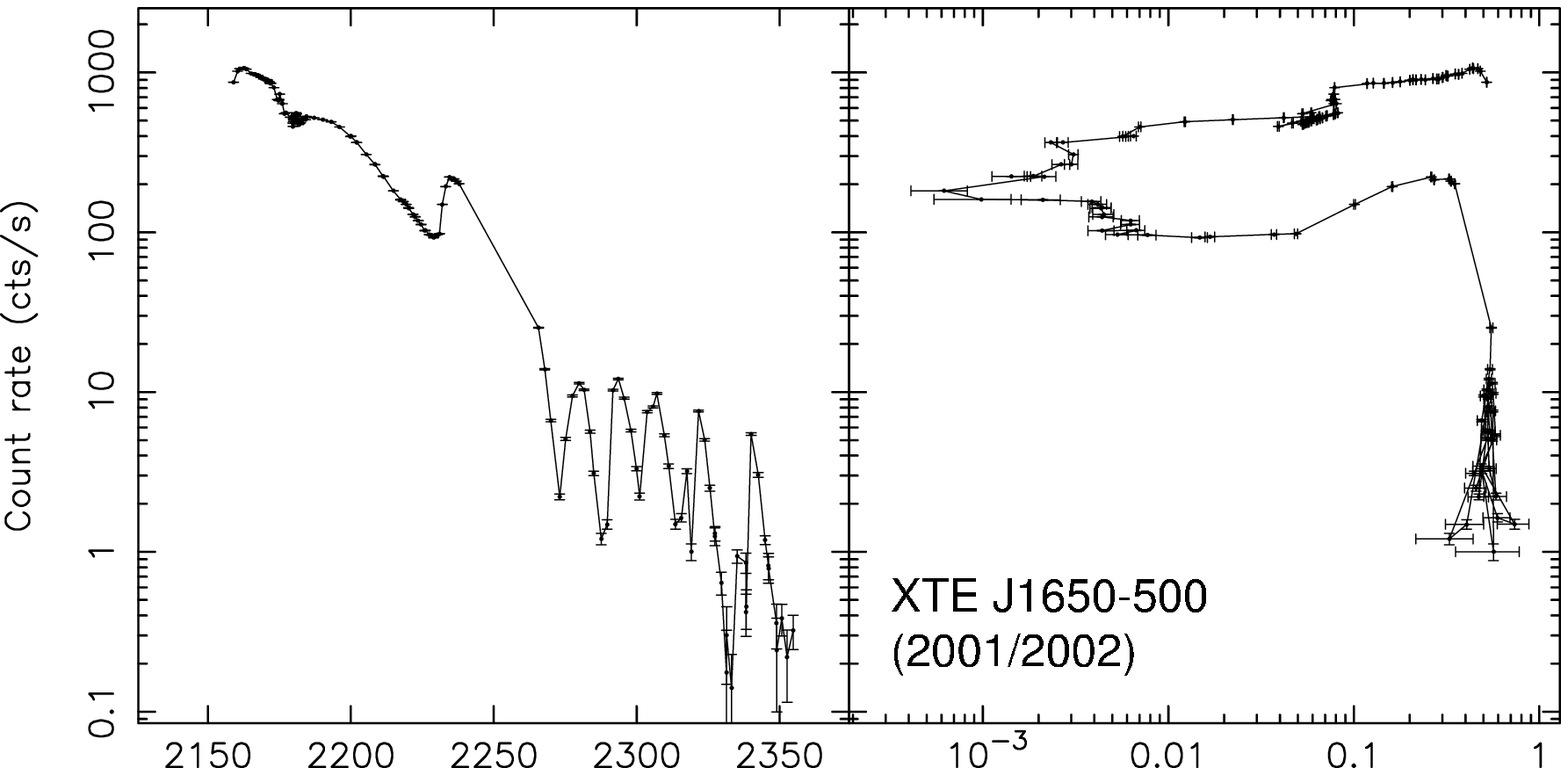}}
\vspace{0.1cm}
\centerline{\includegraphics[width=7cm]{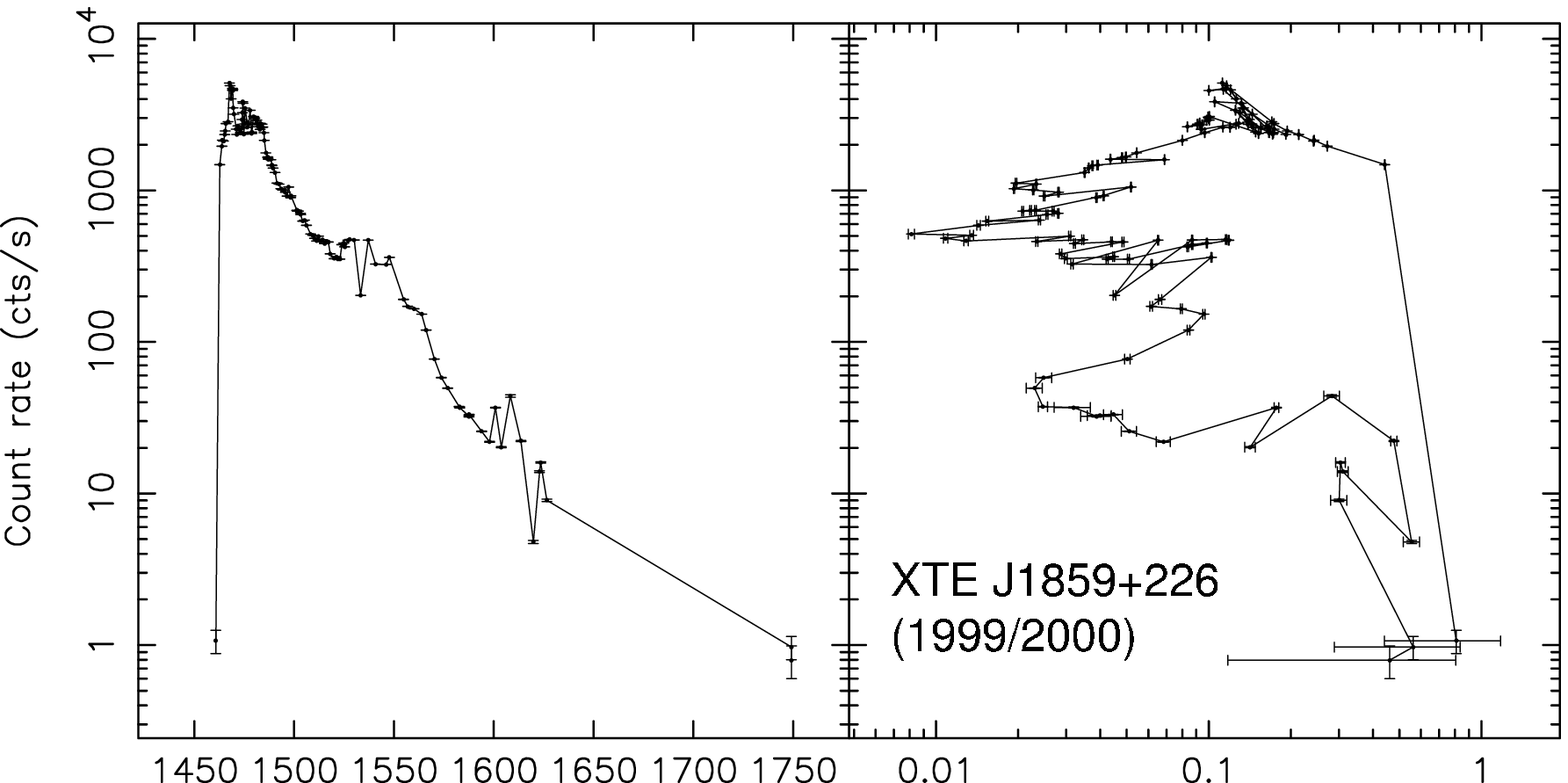}}
\vspace{0.0cm}
\centerline{\includegraphics[width=7cm]{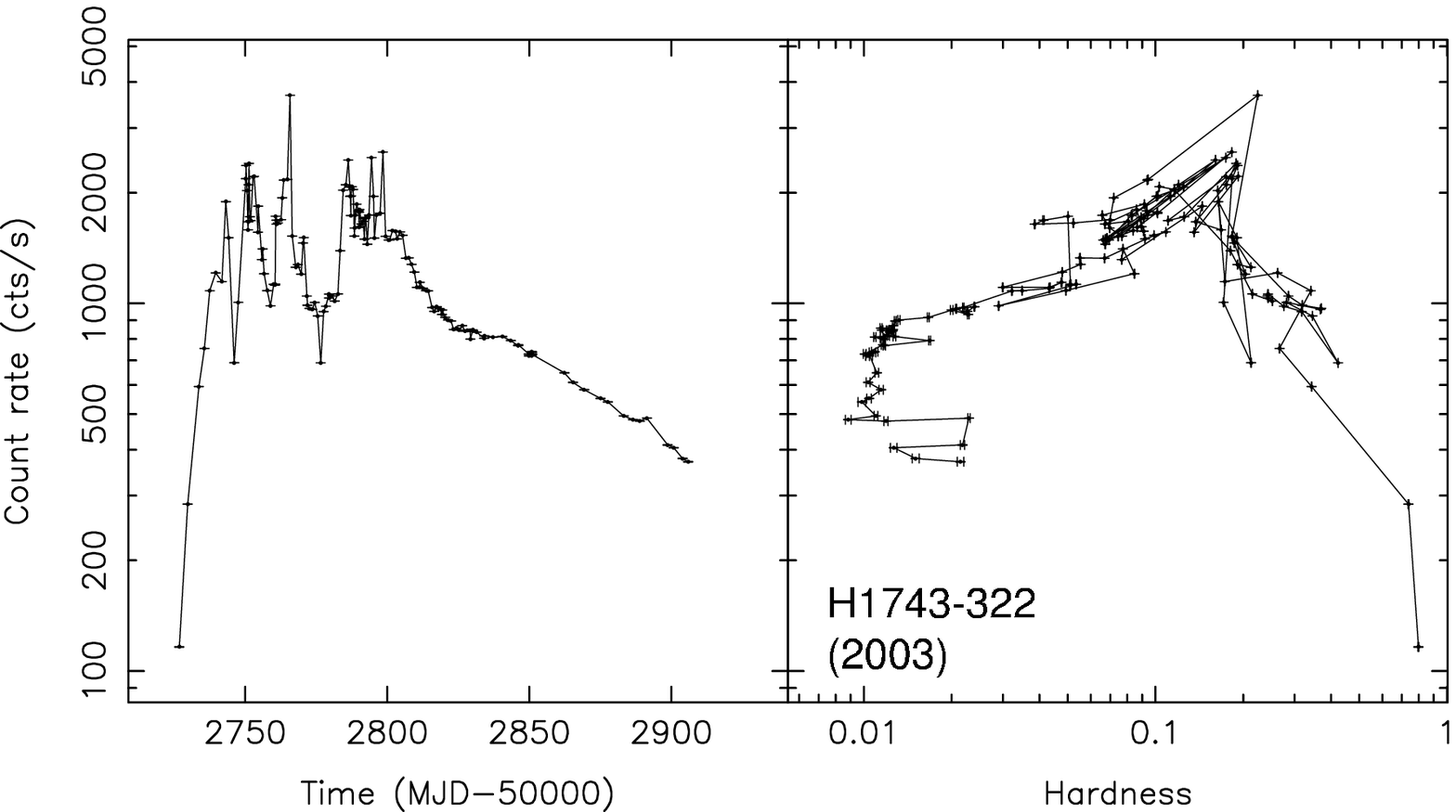}}
\caption{Light curves and hardness-intensity diagrams of four recent 
transients observed with the {\it RXTE}/PCA. Count rates are in the $\sim$3--21
keV band and hardness is defined as the ratio of count rates in the 
$\sim$6--19 keV and $\sim$3--6 keV bands. For GX 339--4, XTE J1859+226, and H1743--322, the outbursts were observed to start in   the upper-right corner, with the sources moving through the diagram in a counter-clockwise direction. For XTE J1650--500 the initial rise was not observed with the PCA, but ASM observations suggest it started in the lower-right corner as well, following a vertical path to the upper-right corner, where the PCA coverage started.}\label{lc-hid}
\end{figure}

Hardness-intensity diagrams (HIDs), in which the X-ray count rate is
plotted  versus an X-ray color, provide a quick way to study the
global evolution of  black hole transients during outburst. In Figure
\ref{lc-hid} we plot the  light curves and HIDs of four transients
that were observed with {\it RXTE}  between 1999 and 2003. It should
be noted that constant count rate levels  in these HIDs correspond to
substantially higher luminosities at hardness values close to 1 than
at values of 0.001-0.01. The HIDs are similar to each other in that 
all four sources seem to trace out (part of) a counter-clockwise
q-shaped  track. Below a certain count rate the spectrum is always
very hard (i.e. in the hard state) and above the spectrum is either
very hard or very soft, except for  two transitional phases (the
'horizontal' branches in the HID). This means, as was noted by other
authors as well (see e.g. \opencite{maco2003}), that within a single
outburst the hard \ra\ soft transition occurs at a flux that is about
a factor of  10--100 higher than the soft \ra\ hard transition. Note that
in GX 339--4 and XTE J1650--500 the hard \ra\ soft transition was
rather fast, while in XTE J1859+226 and H1743--322 the source
lingered for a longer time at intermediate colors.

While the sources shown in Figure \ref{lc-hid} show a smooth overall
movement through their HID,  the fast time variability and
multi-wavelength properties allow one to define a few rather sharp
boundaries corresponding to  state transitions. In the following, we
will base our discussion of states mostly on the 2002/2003 outburst
of GX 339-4, the results of which will presented in \inlinecite{behoca2004} and  Homan et al. (in prep.). Complete references for our
discussion can be found in those works.

\section{A detailed look at the states}

\subsection{The hard state - rise and decay}

\begin{figure}[t] \centerline{\includegraphics[width=7cm]{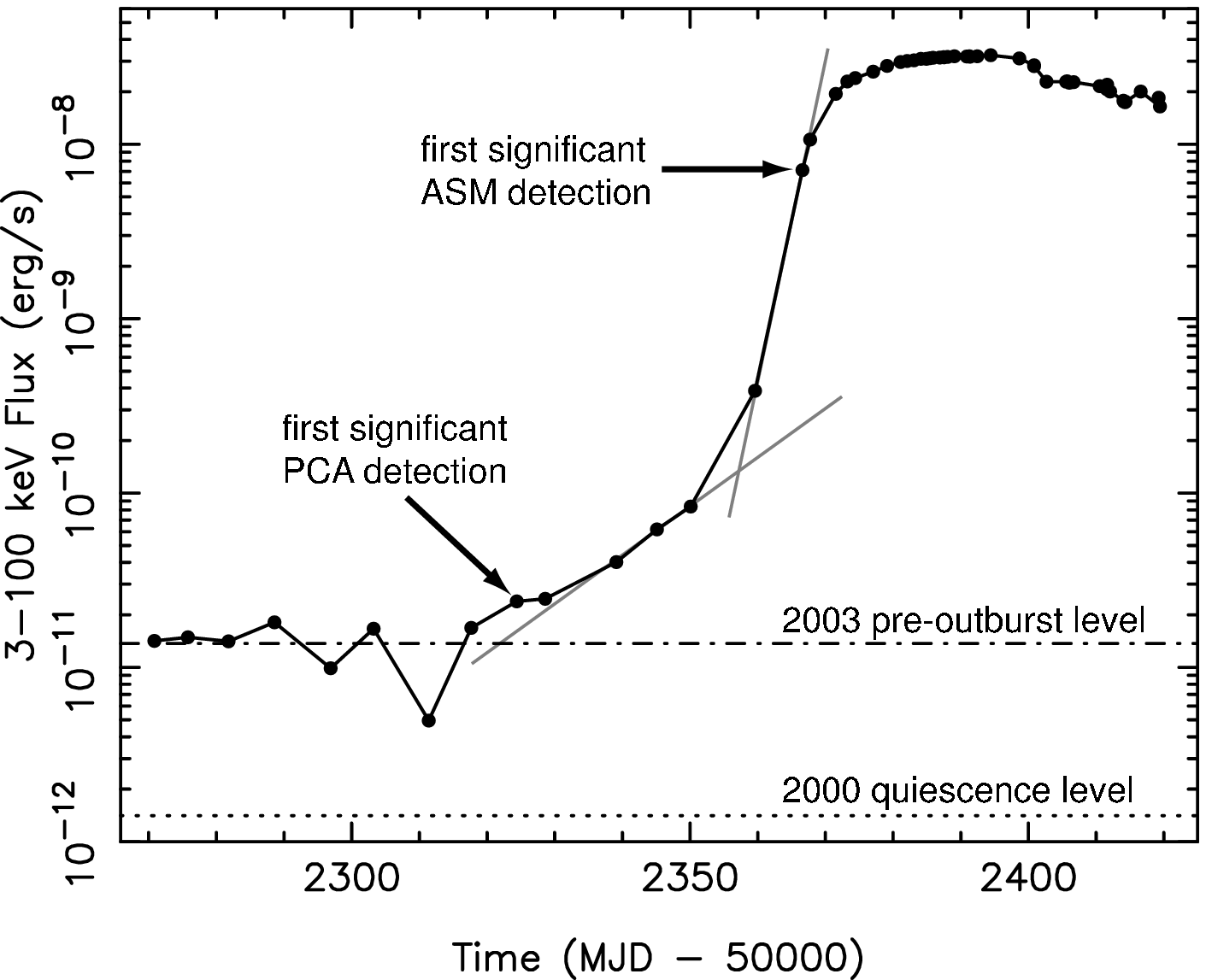}}
\caption{An {\it RXTE} light curve of the rise of the 2002/2003
outburst of  GX 339--4. The times of the first significant PCA and
ASM detections are  indicated, as are the pre-outburst and quiescence
levels. The two gray lines show the presence of two different time
scales during the early rise.} \label{rise} \end{figure}

Since pointed observations of a transient in outburst often start
only when the source already has a luminosity that is a factor of
more than $10^4$--$10^5$ above the quiescent level, not much is known
about the early evolution of outbursts. Thanks to a monitoring
campaign set up by David Smith and  co-workers, the 2002/2003
outburst of of GX 339--4 was the first whose evolution  could be
followed from a flux level that was only a factor of 10 above its 
2000 quiescent level (\opencite{conofe2003}). In Figure \ref{rise} we
show the early evolution of GX 339-4.  From this figure it is clear
that before the first ASM detection the source  luminosity had been
increasing for at least 40 days, initially slowly and later more
rapidly.  

\begin{figure}[t]
\centerline{
\includegraphics[height=8cm]{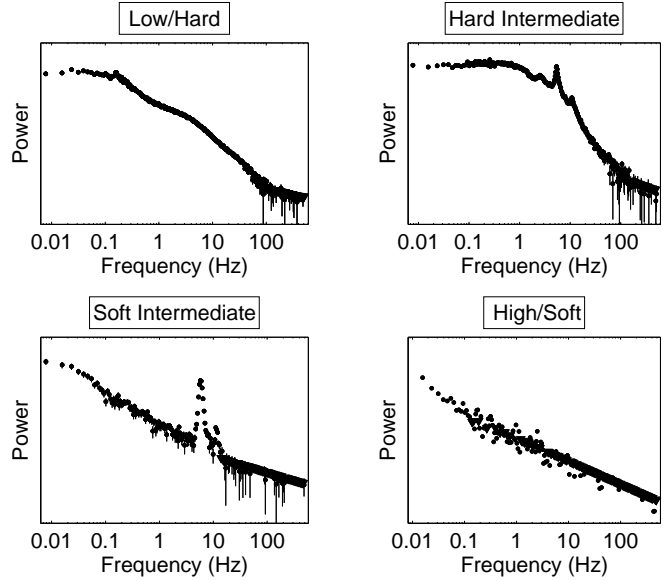}
}
\caption{Examples of power-density spectra of GX 339-4 from the 
four states discussed in the text.
}
\label{pds}
\end{figure}

Throughout these early stages of the outburst, until it had
reached its  peak 3--100 keV flux, the source was in the hard state: 
the strength of the broad-band variability decreased from 45\% rms to
30\% rms and the energy spectrum was dominated by a power law
component, with the index slowly increasing from 1.3 to 1.4. During
the rise the frequencies of the noise components and the occasional 
QPO increased gradually. An example of a power-density spectrum from
this state in GX 339-4 is shown in Figure \ref{pds}.

Just after reaching its peak flux, increases in the power-law index
and  noise/QPO frequencies accelerated. Also, a strong increase of
the spectrally soft disk component and a substantial decrease in the
optical/IR flux (see Fig.~\ref{multiwa}) were observed at that time,
with the latter probably being the result of the jet switching off
(or starting to, \opencite{hobuma2004}).  At this point the source
entered an intermediate  state (see \S \ref{int}), with the
transition from the hard state being visible as an almost 90$^\circ$
turn to the left in the HID. Note that in the definition of
\inlinecite{mcre2003} sources stay in the hard state until the
power-law steepens beyond an index of 2.1. In our view the hard state
is limited to only the hardest power-law dominated spectra with
indices around 1.3--1.5, as many changes occur (simultaneously) when
the power-law becomes steeper.

At the end of the 2002/2003 outburst GX 339-4 returned to the hard
state, similar to what is observed in most, if not all, black hole
transients. In the HID this can be nicely seen as the lower
horizontal branch bending down to start running parallel to, and in
fact nearly on top of, the hard state branch that was traced out
during the rise. Such `saturation' of the spectral hardness was also
observed  in XTE J1650--500 (\opencite{rohomi2003}, see also
Fig.~\ref{lc-hid}). It is interesting to note that once GX 339-4
reached the hardness at which it originally left the hard state
branch, the optical/IR showed a strong increase (see light curves in
\opencite{bafe2004}), indicating that the jet does not become visible
in the optical/IR (or switch on) until the hard state branch is
reached (as was already suggested by \opencite{katobu2004} based on
observations 4U 1543--47). The spectral hardening towards
the hard state is not necessarily a monotonic process, as can be seen
in the HID of GX 339-4 and XTE J1859+226.  

Observations of hard state branches extending over many
orders of magnitude in GX 339-4 and XTE 1650--500 indicate that the
hard state accretion flow geometry (which likely gives rise to strong
outflows) can exist over a wide range in mass accretion rate.

\subsection{The hard state $\leftrightarrow$ soft state transitions}\label{int}

\begin{figure}[t] \centerline{
\includegraphics[height=5.5cm]{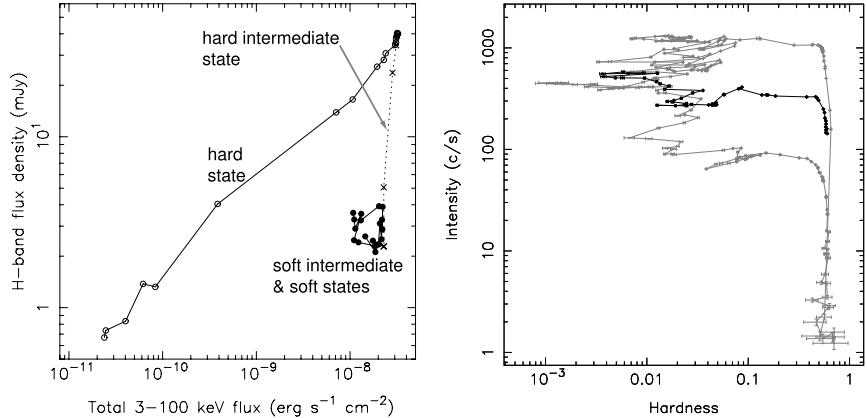} \hspace{0.1cm}
\includegraphics[height=5.5cm]{homan_belloni_f3b.eps}} \caption{Left panel:
X-ray/IR correlation for the first part of the 2002/2003 outburst of
GX 339-4 (Homan et al., 2004). The different behavior for  the
different states is evident. Right panel: HID of the two most recent
outbursts of GX 339-4: 2002/2003 (gray) and 2004 (black).}
\label{multiwa} \end{figure}

Immediately after leaving and before returning to the hard state,
black hole transients are observed in an intermediate state, with
spectral power-law indices between roughly 1.5--2.5. We are not aware
of any exception to this. This state is different from the hard state
in several aspects. Depending on whether the intermediate state is
observed during the hard$\rightarrow$soft or soft$\rightarrow$hard
transition, sources show a clear steepening/\-flattening of the
spectral power law component, a strong increase/decrease in the noise
and QPO frequencies, and an increasing/decreasing fractional disk
flux, compared to the hard state. All these properties evolve
smoothly from and to the hard state. In fact, without the HID and
optical/IR information (see Fig.~\ref{multiwa}) it would be difficult
to exactly pin-point the transition between these two states.  In the
case of the hard\ra soft transition the movement to the left in the
HID is due to the combination of increased disk flux and steepening
of the power law.  In the timing domain, the broad-band variability
components seen in the power density  spectrum increase their
characteristic frequencies, showing an evolution that clearly links
them to the corresponding components in the hard state. A clear
type C QPO appears, also with a characteristic
frequency increasing with time and decreasing hardness (see
\inlinecite{behoca2004}). A typical power-density spectrum from the
intermediate state is shown in Figure \ref{pds}.

It is important to note that the hard$\rightarrow$intermediate state
transition does not always occur at the  same flux level, as can be
seen from Figure \ref{multiwa}.  During the 2004 outburst of GX
339--4 this transition occurred at a flux level that was about a
factor of 4 lower than in 2002. The transition in 2004 was preceded
by a small hard state outburst, which suggests that the flux
level at which the transition  occurs depends on the recent accretion
history. After passing through the intermediate GX 339-4 continued to
brighten in the soft sate during its 2004 outburst.

Another important point to note is the fact that the occurrence of
the intermediate  state  does not seem to depend on  the time
derivative of $\dot{M}$. In XTE J1650--500 it occurred during the
decay, but in GX 339--4 while $\dot{M}$ was apparently still
increasing,  as witnessed by the second (soft) maximum in the light
curve. However, in  both cases the transition resulted in a
(temporary) drop in the count rate  by a factor of $\sim$2.

Taking once more the 2002/2003 outburst of GX 339--4 as a template,
we see that when the hardness goes below a well defined threshold,
the timing properties change sharply: a clear type B QPO appears in
the power density spectrum (Figure \ref{pds}, see also
\opencite{cabeho2004} for similar behavior in XTE J1859+226). In most
sources the change from type C to type B QPOs seems to take place
when the power-law index has a value around 2.5-3.0. In fact, power-laws with such indices are accompanied by a great variety of timing properties: not only type A, B or C QPOs, but also very weak variability like that seen in the soft state.

We will refer to the part of the transition during which the
power-law index changes between $\sim$1.5--2.5 and which shows type C
QPOs and strong band-limited noise as the hard intermediate state.
The part of the transition where the power-law index is relatively
constant around a value of 2.5--3.0 and during which the source
occasionally shows type A and B QPOs on top of weaker red noise will
be referred to as the soft intermediate state.  Note that in GX 339-4
the soft intermediate state also showed observations with weak band
limited noise and/or QPOs that we were not able to classify. It is
therefore more a collection of different types of behavior, between
which the source could switch on a time scale of a day, rather than a
well defined state.

It is important to note that the change from the hard intermediate to
the soft intermediate state was repeated again in the 2004 outburst
at the same spectral hardness. During the  2002/2003  outburst a huge
radio flare/ejection event was observed around the time of the change
from the hard intermediate to the soft intermediate state
(\opencite{gacofe2004}, \opencite{febega2004}). Although  it is
tempting to associate such a flare with this transition, it should 
be mentioned that similar radio events are also observed in GRS
1915+105 (\opencite{beklme2000}),  a source for which characteristic
type A and B QPOs were not observed to date. However, given the fast
time scales of transitions in this system, it is possible that such
QPOs appear for intervals shorter than for other sources, which would
make them difficult to observe.

It is during the soft intermediate state that most high-frequency
QPOs have been detected, indicating that there might be a relation
between those and type A/B QPOs;  indeed, the frequencies of both
features are not observed to vary by a large amount between different
observations. It is possible however that the high frequency QPOs in
soft intermediate state evolve from broader features in the hard
intermediate state, as is suggested by observations of XTE J1650--500
(\opencite{hoklro2003}). Notice that while the soft intermediate
state is often observed at high flux between the hard intermediate
and the soft states, the same timing properties are not observed at
low flux, when the reverse transition takes place. Moreover, in XTE
J1550--564 (Homan et al., 2001) type A and B QPOs  appeared at several
well separated luminosity levels, strongly suggesting they are not
strictly related to hard$\leftrightarrow$soft transitions. In GX
339-4 a strong 1 Hz QPO with some properties similar to those of the
type B QPOs is observed at low flux, indicating that it is possible
that a similar state exist also at much lower accretion rate,
although with different characteristic frequencies
(\opencite{behoca2004}).

\subsection{The soft state}

During the soft state, which in GX 339-4 took place after the soft
intermediate state, the spectral and timing properties are rather
well defined, although a single clear transition from the soft
intermediate state is hard to identify. The energy spectrum of the
soft state is dominated by a strong thermal component, with the
presence of a weak steep power-law component, which was not observed
to show a high-energy cutoff (see Grove et al., 1998).  Variability in
the soft state is weak compared to the other states, with typical rms
amplitudes of at most a few \% rms. Unfortunately, by the time most
transients reach the soft state they are usually in the decay phase
and observations have become shorter and less frequent, so detailed
studies of its variability properties are rare. Nevertheless, a few
weak QPOs have been detected in the soft states of GRO J1655--40, XTE
J1550--564, and H1743--322. These QPOs all had frequencies that were
higher than the other low frequency QPOs detected in those sources
and the noise continua could all be fitted with broken power law.
Surprisingly, the QPO and break frequencies of these power spectra
fall on top of the Wijnands-van der Klis relation (Wijnands \& van
der Klis 1999) for black holes in the hard and hard intermediate
states. This  suggests not only that these might be type C QPOs, but
more importantly, that variability properties that once were thought
to be characteristic of the hard and hard intermediate states are
still present in the soft state, although in a much weaker form.

\begin{figure}[t] \centerline{
\includegraphics[height=8cm]{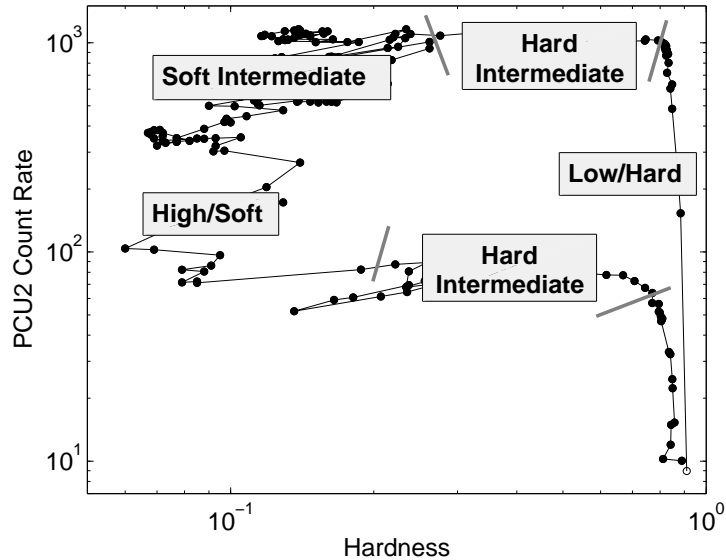} } \caption{The HID
of the 2002/2003 outburst of GX 339-4. Clear transitions are marked
by gray segments. The branches corresponding to the four basic states
in the q-track are labeled (Belloni et al., 2004).} \label{hid339}
\end{figure}

\subsection{Discussion}

Although the picture presented here is somewhat simplified, as
additional complications have been observed, we can sketch the states
by using once more GX 339-4 as a template (see Fig. \ref{hid339} and
\opencite{febega2004}).  As shown above, the hard intermediate and
soft intermediate states are kept separate, as both spectral and
timing evolution show marked differences. 

The hard state and the hard intermediate states have much in common.
Spectrally, they are dominated by a hard component for which a
high-energy cutoff is observed (\opencite{grjokr1998}).  In the
timing domain, the components observed in the power density spectrum
in these two states are clearly related, as can be seen from the
evolution of their characteristic frequencies
(\opencite{behoca2004}). Nevertheless, something happens to the
accretion flow as the source moves from the low/hard state to the
intermediate state, as can be seen from multi-wavelength studies
(\opencite{hobuma2004}). This is likely to be related to changes in
the jet, which could be responsible for part of the observed X-ray
flux (\opencite{febega2004}) as is also indicated by the fact that
radio emission is always observed in these states.

The soft state and soft intermediate state are spectrally somewhat
related. The energy spectrum is dominated by the thermal disk
component, with a steep hard component with no evidence of a
high-energy cutoff. The power density  spectrum lacks band-limited
components and shows only QPOs superimposed on  a power-law
component. No radio emission is observed (i.e. only upper limits on
the emission from the compact source), indicating that the production
of a jet is terminated at the transition to  soft intermediate state
(\opencite{febega2004}) and maybe even before. During these states,
the accretion flow is clearly different from the other two. In terms
of variability however, the  soft state  shares some properties with
the hard state and hard intermediate state, with the typical
variability time scales following the same relation as seen in those
states. If the hard state variability properties are  linked to jet
production, the presence of related (but much weaker) variability in
the soft state could indicate a very weak (i.e. below current
detection limits) jet in that state as well.

In addition to these general properties, there is a number of
important topics whose detailed discussion is beyond the scope of the
present paper.  Most notably, the fast transitions observed between
different states (never involving the hard state, which is only
reserved for the beginning and the end of an outburst) need to be
studied in detail, as they probably hold the key for a deeper
understanding of the physics of the states and their association to
the jet. Also, some sources can have bright outbursts without ever
leaving the hard state, indicating that once again the instantaneous
mass accretion rate is not what determines the transitions. However,
the recent history of the mass accretion rate may play an important
role (this can also be seen in Fig. \ref{multiwa}, as GX 339-4 was
observed to leave the hard state are very different flux levels). A
secondary accretion flow (\opencite{smhesw2002};
\opencite{yuvafe2004}) or changes in the Compton cooling/heating in
the accretion disk corona (\opencite{melime2004}) seem to be
promising to explain some of these issues. For example, delays
between a fast (i.e. in terms of propagation speed) secondary flow
and a slower disk flow could account for the observed hysteresis.  

In conclusion, the hysteresis behavior that can be seen from the HIDs
in Figure \ref{lc-hid}, put forward for the first time by
\inlinecite{mikiig1995}, was found with {\it RXTE} to be a
characteristic pattern for black-hole transients. Despite many
complexities, general features have been shown to relate to the
ejection of powerful relativistic jets, which can lead to a deeper
understanding of the  physical properties of the accretion flow. The
picture presented here, scaled to longer time scales, should  also be
valid for AGNs. In these sources, there is no contribution of the
optically thick accretion disk in the X-ray band, but the path shown
in Fig. \ref{hid339} is qualitatively similar if the disk
contribution is removed. Recently, \inlinecite{cu2004} presented a
very similar diagram for the {\it RXTE} observations of the highly
variable AGN Mkn 421, showing that indeed the scheme of accretion
states could apply to systems over a large scales of masses.
\\

\noindent JH wishes to thank Jon Miller and Ron Remillard for numerous discussions on the topic of black-hole states.


\end{article}

\end{document}